\documentclass[a4paper,12pt]{article}

\usepackage[utf8]{inputenc}
\usepackage[english]{babel} 

\usepackage{graphicx,epsfig}

\usepackage{amsmath,amssymb,amsfonts,amssymb}

\usepackage{authblk}
\begin{document}

\title{Default contagion risks in Russian interbank market}
\author[1,2,3,4]{A.V. Leonidov \footnote{Also at ITEP, Moscow}}
\author[2,5]{E.L. Rumyantsev \footnote{\indent The material presented does not necessarily reflect the position of the Bank of Russia on the issues under discussion. }}
\affil[1]{Theoretical Physics Department, P.N. Lebedev Physical Institute, Moscow}
\affil[2]{Chair of Discrete Mathematics, Moscow Institute of Physics and Technology} 
\affil[3]{Center for the Study of New Media \& Society, New Economic School, Moscow}
\affil[4]{Laboratory of Social Analysis, Russian Endowment for Education and Science, Moscow} 
\affil[5]{Department of Financial Stability, Bank of Russia, Moscow}

\date{}

\maketitle

\renewcommand{\abstractname}{Abstract}

\begin{abstract}
Systemic risks of default contagion in the Russian interbank market are investigated. The analysis is based on considering the bow-tie structure of the weighted oriented graph describing the structure of the interbank loans.  A probabilistic model of interbank contagion explicitly taking into account the empirical bow-tie structure reflecting functionality of the corresponding nodes (borrowers, lenders, borrowers and lenders simultaneously), degree distributions and disassortativity of the interbank network under consideration based on empirical data is developed.  The characteristics of contagion-related systemic risk calculated with this model are shown to be in agreement with those of explicit stress tests.
\end{abstract}

\newpage

The recent financial crisis has brought into the focus of attention systemic risks related to economic interactions between banks. Such interactions are most naturally described in terms of a network of their mutual obligations \cite{IMF_2009,Haldane_2009,BIS_2012a,BIS_2012b}. The notion of systemic risks refers to various crisis phenomena involving many economic agents having their origin in their interaction, e.g. in the outstanding interbank loans \cite{May_2011,Sornette_2011,Moiseev_2012,Chinazzi_2013,Leonidov_2014}. Game-theoretical foundations for building a quantitative description of financial contagion were discussed in Refs.~\cite{Acemoglu_2013_a,Acemoglu_2013_b,Eliott_2013}.

One of the most important types of crisis phenomena taking place on complex networks are epidemic type cascading processes, see Ref.~\cite{Borge_2013}. In the particular case of interbank network considered in the present paper this is a contagion process triggered by the default of some bank possibly followed by defaulting of some of its neighbors and, finally, to a formation of a default cluster. The phenomenon of contagion in interbank markets has drawn a lot of attention in the literature, see e.g. the recent reviews \cite{Chinazzi_2013,Leonidov_2014}. Foundations of quantitative analysis of contagion spreading in interbank networks were laid in the paper \cite{Kapadia_2010}\footnote{See also an interesting comment \cite{Lo_2011}.}, in which the underlying interbank network was considered as a weighted directed Poissonian random graph with neither clustering nor degree-degree correlations taken into account. The focus of \cite{Kapadia_2010} was on systemic risk associated with the percolation phase transition and formation of giant cluster and the related robust-yet-fragile property of default contagion where a small probability of a catastrophic event goes together with its huge volume. 

The research following  \cite{Kapadia_2010} was to a large extent aimed at checking the assumptions on the properties of interbank network made in \cite{Kapadia_2010} against existing data and, if necessary, incorporating the corresponding modifications into the theoretical formalism describing default propagation. In terms of network properties the question is thus on degree distributions, degree-degree correlations and clustering in realistic interbank networks, while theoretically a central issue is that of an interplay between topological properties of a network and those of epidemic cascades on it.

Analysis of empirical properties of the Russian interbank network was made in Refs.~\cite{Leonidov_2012,Leonidov_2013} for the period of  01.08.2011-03.11.2011 and in Refs.~\cite{Vandermaliere_2012,Vandermaliere_2015} for the period of 10.2004-08.2008. It was found that this network is characterized by a heavy-tailed degree distribution \cite{Leonidov_2012,Leonidov_2013,Vandermaliere_2012,Vandermaliere_2015}, heavy-tailed distribution of exposures  \cite{Vandermaliere_2012,Vandermaliere_2015}, pronounced disassortative degree-degree correlations and significant anomalous clustering  \cite{Leonidov_2012,Leonidov_2013} \footnote{Similar results for degree distribution, degree-degree correlations and clustering were found for the Brasilian interbank market in \cite{Santos_2010}.}. In \cite{Caccioli_2012} effects of taking into account heavy-tailed nature of degree distributions, degree-degree correlations and clustering on default contagion propagation in the framework of \cite{Kapadia_2010} were studied, on a feature-by-feature basis, in a Monte-Carlo simulation. The first Monte-Carlo simulation on interbank network with realistic topology was performed in \cite{Leonidov_2012,Leonidov_2013}.

On theory side it is necessary to incorporate the above-mentioned characteristic features of real interbank networks (heavy-tailed degree distributions, disassortativity and anomalous clustering) into an analytical formalism. The generating function formalism used in \cite{Kapadia_2010} is directly applicable for arbitrary degree distributions, so the remaining problem is to take into account degree-degree correlations and clustering. There exists a considerable literature on the effects of the influence of these features on percolation transition and epidemic diffusion. Effects of degree-degree correlations on the percolation transition was studied in \cite{Newman_2002_b} for undirected and \cite{Boguna_2005,Gleeson_2008} for directed graphs, while that of clustering was analyzed in Refs.~\cite{Hackett_2011,Melnik_2011}.

The present paper continues investigation of systemic risks at the Russian interbank market begun in Refs.~\cite{Leonidov_2012,Leonidov_2013} and focuses on developing a mathematical model of contagion process taking into account all aspects of geometry of interbank network, important features of the balance sheet structure of participating banks and institutional regulation relevant for providing a quantitatively correct description of default cascades. We argue that to build such a description one should take into account empirical default propagation probability, bow-tie structure, degree distribution and disassortative correlation structure of interbank networks. An interesting feature we observe is that although the original interbank network is characterized by high clustering, default clusters are predominantly tree-like. This latter property agrees with the findings of Refs.~\cite{Melnik_2011} and \cite{Goel_2013}\footnote{We are grateful to C. Borgs for pointing out this reference.}.  

The main objectives of this paper are the development of the above-described analytical model based on an enlarged set of empirical data on the Russian interbank market. 

The plan of the paper is as follows.

In section~\ref{section:empirics} we discuss the structure and main characteristics of the Russian money market and the data used in the analysis. In the section \ref{section:deposit} we analyze empirical characteristics of the deposit market from the network perspective including its bow-tie structure, degree distributions and correlations, clustering and default propagation probabilities.  In section \ref{section:risk_empirics} we explore the structure of the default cluster caused by the default of a randomly chosen bank and after that introduce mathematical formalism for systemic risk representation in chapter \ref{section:risk_theory}. Conclusions are presented in section \ref{section:conclusions}.

\section{Russian money market: empirics and data description}
\label{section:empirics}

Russian money market consists of three main segments, the markets of deposits, REPO and SWAP. 

Operations at the deposits market are uncollateralized: banks bound their risk in lending money to counterparties by setting limits calculated with the help of in-house models and taking into account expert opinions. Lending risks are also regulated by the special requirement of the Bank of Russia constraining the value of exposure to a counterparty. The uncollaterized nature of the deposit market makes it the most vulnerable with respect to trust evaporation during crises when deposit markets can freeze requiring significant efforts from regulators for their relaunching.  

Less risky is the REPO market at which collateral, usually government and corporate bonds and equities, is required.  In operations with money borrowing the value of different types of collateral includes discounts thus reducing the corresponding market risk. Credit risk is more often accounted for in the credit rate. Let us note that while at the beginning of the Russian crisis in 2008 the REPO market in Russia did collapse, it started functioning faster than the deposit one\footnote{The collapse of deposit and REPO market took place after the Lehman Brothers default. Due to efforts of Bank of Russia and Ministry of Finance the Repo market was partly retrieved in two weeks whereas the credit risks still remained too high.}.

The least risky is the SWAP market. Often swap operations are used as a source of short-term ruble liquidity when in exchange of rubles a lender gets foreign collateral (most often USD and EUR). SWAP operations were attractive for Russian banks during the period of systemic liquidity deficit when banks are permanently in need of liquidity refinancing from central bank. 

We provide comparative statistics for outstanding in different segments of the Russian money market in Table \ref{T_Market_outstanding}. For deposit market it includes claims on banks (resident and non-resident) in deals with Russian rubles. For SWAP market it includes claims on resident and non-resident banks in rubles against USD and euro as collateral. It is worth saying that the data includes deals between banks at the OTC market as well as  through Central Counterparty (CCP), the latter option was used in almost 25\% of swap deals. The Repo market data includes claims on banks in rubles. In 2013  the project of REPO deals through Central Counterparty was launched. The total outstanding on CCP at the end of 2013 year was 1.5 bln. USD. To highlight the importance of money market we also provide the total value of assets of the Russian banking system. Some information concerning evolution of the outstanding for different money market segments can be found in the Bank of Russia ''Money market report'' \cite{CBRF_market}. 


\begin{table}\label{T_Market_outstanding}
\begin{center}
\begin{tabular}{|c|c|c|c|}
  \hline
  Market $\backslash$ Year  & 2011 & 2012 & 2013 \\ \hline
  Deposit & 52.6 & 54.1 & 55.0 \\ \hline
  REPO & 5.4 & 6.4 & 5.2  \\ \hline
  SWAP & 94.5 & 84.4 & 101.2  \\ \hline
  Total assets& 1318.9 & 1609.6 & 1742.6  \\ \hline
\end{tabular}
\caption{Volumes of deposit, REPO and SWAP markets and total market of banking activities in Russia in Bln. USD \cite{CBRF_market}. Total assets of banking system are given to compare the volume of money market with that of other banking activities such as corporate and retail lending, investments in securities, and others.}
\end{center}
\end{table}

We based our analysis on daily CBRF banking report "Operations on currency and money markets" \cite{CBRF_report} containing information on all type of transactions carried out on the OTC money market \footnote{A decentralized market, without a central physical location, where market participants trade with each other through various communication modes such as telephone, email and proprietary electronic trading systems. An over-the-counter (OTC) market and an exchange market are the two basic ways of organizing financial markets.}. In our analysis we concentrate on the deposit market and take into account only  uncolleteralised deals in Russian rubles between residents. Taking into account only the deals between residents is due to data limitations. We also exclude deals with Central Bank of Russia and its branches because corresponding ruble obligations are always met and therefore do not generate any risk. 

Analysis of other segments of the interbank market goes beyond the scope of this paper. For an interesting analysis of the properties of the multilayer network including  REPO, SWAP and deposit interbank markets in Italy see Ref.~\cite{Borgigli_2015} where it was found that all segments of interbank market share such common features as fat-tailed degree distributions, disassortativity and large values of clustering coefficients. In terms of contagion we would expect that taking into account other money market segments may significantly influence the corresponding exposures and amplify the volume of contagion.

The data used in our analysis cover the interval from January 11 2011 till December 30 2013 and contain information on interbank loans to residents of 185 banks. This data corresponds to roughly 80 \% of the total outstanding and can therefore be considered as representative. 

\begin{figure}[h!]
\centering
\includegraphics[width=0.9\linewidth]{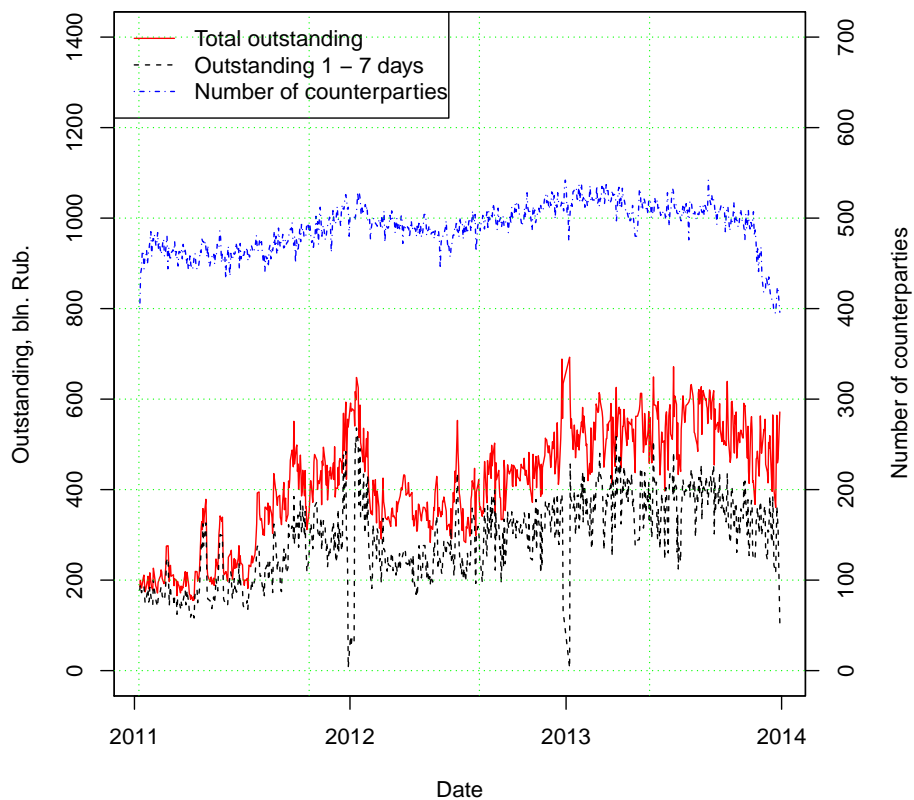}
\caption{Deposit market term structure outstanding}
\label{FOutstandingCounterparties}
\end{figure}

Having information on interbank transaction we transformed it into outstanding for each bank with respect to all of its counterparties on the daily basis. Claims and obligations with different maturities were just summed up. From the credit risk point of view the maturity of the deal is not important if a counterparty has already defaulted on its obligations. Of relevance is then only  the value of claims on such a counterparty.
As our data cover time interval from January 11 2011 till December 30 2013  we have no information on loans borrowed before January 11 January and returned afterwards. It is not important for short term maturity operations but may be important for large maturities. From Fig.\ref{FOutstandingCounterparties} showing dynamics of outstanding for deposit operations we see that deals with maturities 1-7 days constitute a significant part of the total outstanding. Long term maturity deals form about 30\% of the total outstanding. The number of counterparties having at least one deal at a given day slightly changes with time sharply declining at the end of 2013 due to actions Bank of Russia aimed rising banking solvency. 

\section{Interbank deposit market: structure and properties}
\label{section:deposit}

For the purposes of our analysis we view the interbank credit market as a weighted oriented graph characterized by the weighted adjacency matrix $W=\{ w_{ij} \geq 0 \}$ where link variables $w_{ij}>0$ correspond to liabilities of the bank $i$ towards the bank $j$ and are computed by netting the mutual obligations of both banks on the daily basis so that a directed link $i \to j$ corresponds to a credit to $i$ provided by $j$. For a given node outgoing (out-type) links correspond therefore to its obligations towards neighboring nodes and incoming (in-type) ones to claims of the node under consideration towards neighboring nodes. Let us note that with this choice of notations default cascades triggered by some initial node propagate along link's direction.

\subsection{Bow-tie decomposition}   

In describing systemic risks related to network topology of the interbank market it is essential to take into account the gross structure of the corresponding oriented graph represented by its bow-tie decomposition, see e.g. \cite{Newman}. In the problem under consideration, following  the above-described definition of the weighted adjacency matrix $W$, the bow-tie structure separates, on the daily basis, the nodes (banks) into four groups according to the type of their operations. The Out- and In- components include nodes having only outgoing and incoming links correspondingly, i.e. include pure borrowers and lenders respectively. The In-Out- component includes banks having both incoming and outgoing links which are therefore both creditors and lenders. We will show that banks belonging to this component play a crucial role in generating systemic risks. The last group consists of nodes without links. The bow-tie structure includes as a particular important case the core-periphery model with its core belonging to the In-Out- component and periphery to the In- and Out-  ones. 



The structural analysis of the data on interbank network shows that most of the banks  having links (60\%-70\%) belong to the In- component, i.e. act as pure lenders while only 10\%-20\% belong to the Out- component and act as pure borrowers. The number of pure borrowers and lenders displays pronounced seasonality so that the number of the former increases and of the latter decreases at the beginning of the year. As to the structure of the outstanding, it is predominantly concentrated in the In-Out- component (60-70\% of total outstanding) so that the the corresponding banks have a persistent tendency of borrowing (lending) within the In-Out- component. Less important from exposure point of view are the links between the In-Out and In components (20-35\% of total outstanding) and the Out and In-Out ones (2-20\% of total outstanding). The links between the pure borrowers and pure lenders contain less than 5\% of the total outstanding. The In-Out component contains a strongly connected one (SCC) with the size varying from 10\% to 15\% and carrying 40-60\% of the total outstanding demonstrating a significant monopolistic power of several influential actors. 

\subsection{Quantitative characteristics}

In this paragraph we discuss some most important quantitative characteristics of the Russian interbank network such as distributions in the number if incoming and outgoing links, correlations in the degrees of neighboring nodes and degree of clustering.  

\begin{figure}[h!]
\begin{minipage}[h]{0.45\linewidth}
\centering
\includegraphics[width=\textwidth]{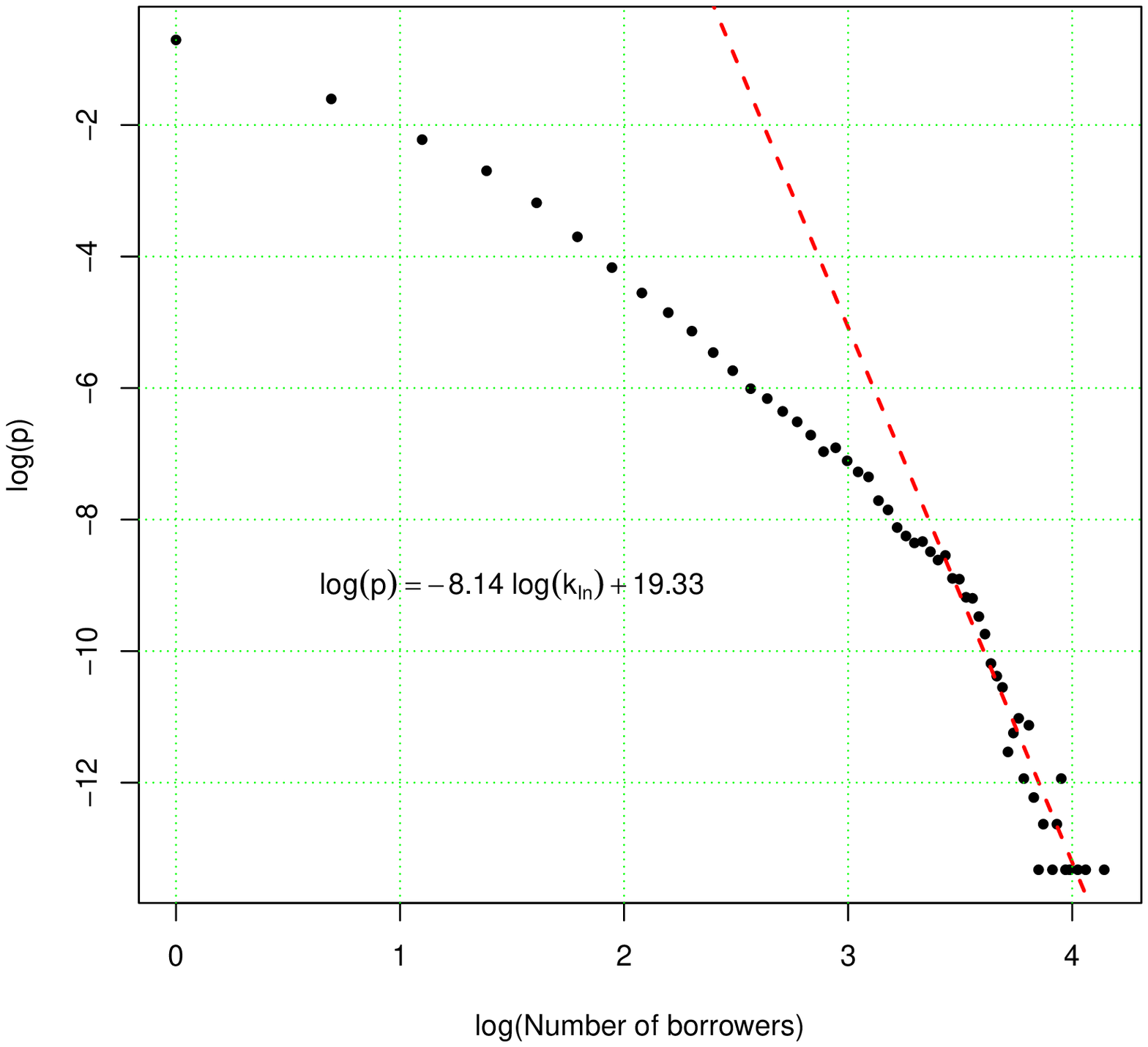}
\caption{Number of borrowers distribution}
\label{FInDegreeDistr}
\end{minipage}
\hspace{0.5cm}
\begin{minipage}[h]{0.45\linewidth}
\centering
\includegraphics[width=\textwidth]{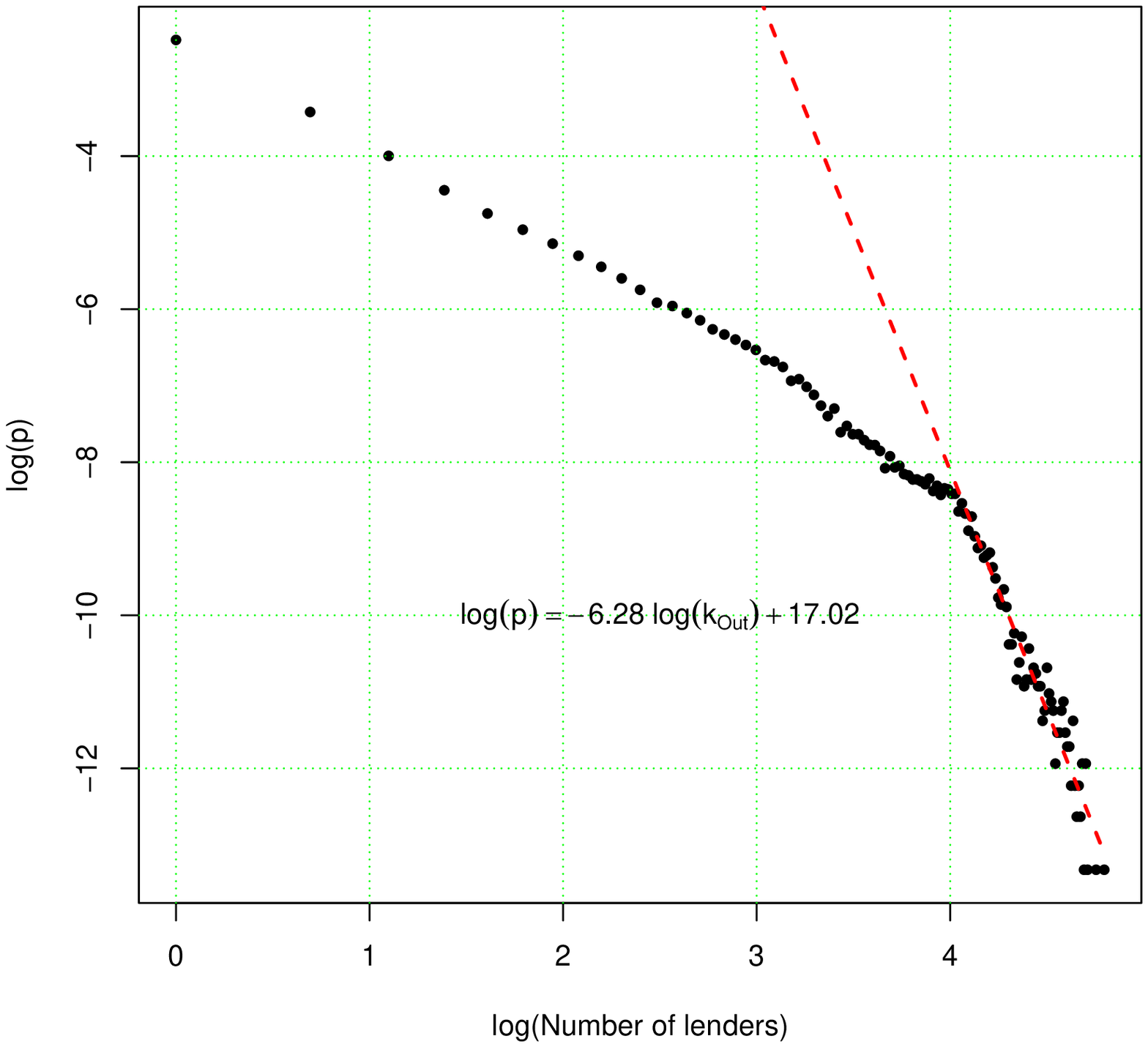}
\caption{Number of lenders distribution}
\label{FOutDegreeDistr}
\end{minipage}
\end{figure}

The distributions of in- and out- degrees are plotted in Figs. \ref{FInDegreeDistr} and \ref{FOutDegreeDistr}. We see that these distributions are fat-tailed\footnote{The small number of observations in the fat-tailed part of the distributions doesn't allow us to test a hypothesis on its power-law nature. According to  \cite{Clauset_2009} the sufficient number of observations is around 1000 while we have only 50-60.} . This feature is in agreement with other results on interbank networks in the literature, see e.g.  \cite{Santos_2010}  and references therein.  

\begin{figure}[h!]
\centering
\includegraphics[height=0.3\textheight,width=0.45\linewidth]{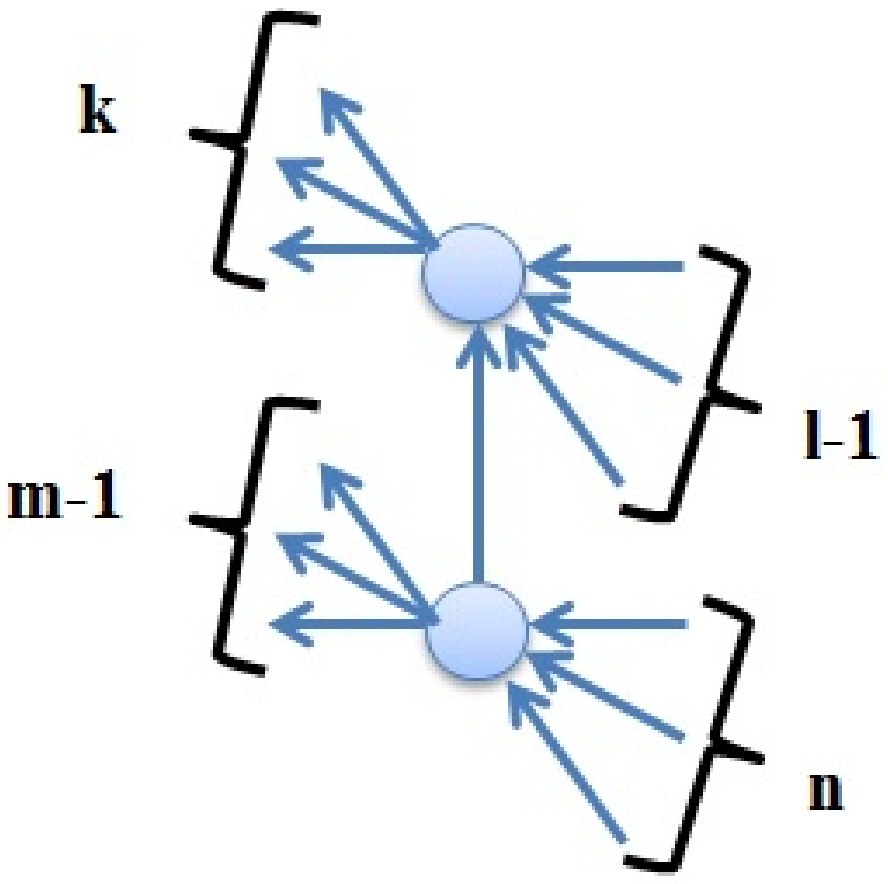}
\caption{Bivariate distribution $P^{\rm IO \to IO}(k,l \vert m,n)$}
\label{PAB}
\end{figure}

An analysis shows, in agreement with the results of \cite{Santos_2010}, that Russian interbank network is characterized by pronounced negative degree correlations (disassortativity). In oriented graphs one deals with several types of degree correlations induced by the bow-tie structure of the network. Probabilistic interdependencies of degrees of adjacent nodes are fully characterized by the set of bivariate distributions $P^{\rm A \to B}(k,l \vert m,n)$, where $A,B$ refer to the component of the bow-tie decomposition $\rm \{I,IO,O\}$ denoting In, In-Out and Out components respectively and the indices $k,l$ and $m,n$ denote the in- and out-degrees of the adjacent vertices, see Fig.~\ref{PAB} in which $P^{\rm IO \to IO}(k,l \vert m,n)$ is shown. 
In section \ref{section:risk_empirics} we will see that in order to provide a good description of the empirical systemic risk characteristics one has to take into account the probability patterns described by $P^{\rm IO \to IO}(k,l  \vert m,n)$, etc.    

\begin{figure}[h!]
\centering
\includegraphics[height=0.45\textheight,width=0.85\linewidth]{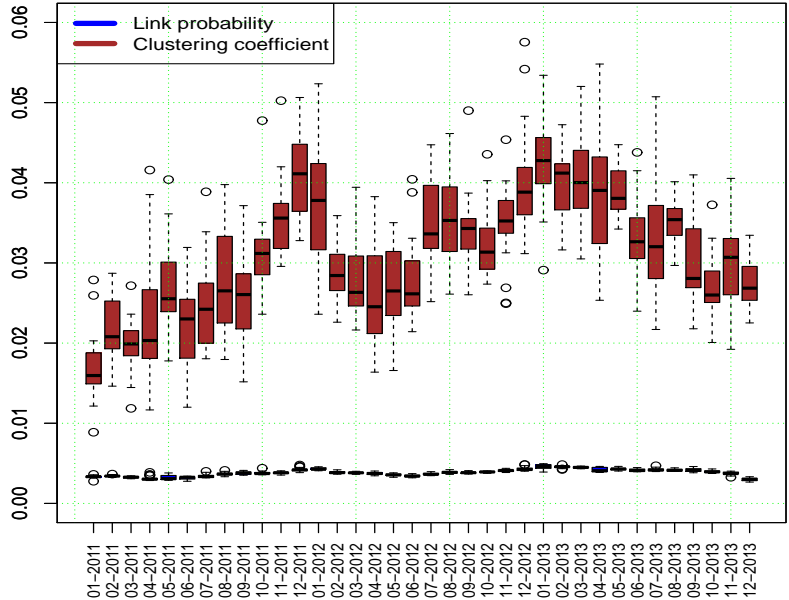}
\caption{Evolution of clustering coefficient and link probability}
\label{clust}
\end{figure}

The interbank network is characterized by significant clustering. This feature is illustrated in Fig.~\ref{clust}, in which time series for clustering coefficient\footnote{Here for simplicity we consider the graph under consideration as unoriented.} and link probability (defined as the ratio the number of links in network to the maximal possible number of links in network)  are shown. We see that the graph under consideration is significantly less sparse than the Erdos-Renyi one fully specified by the link probability for which the clustering coefficient is simply equal to link probability.

\section{Systemic credit risk. Empirical analysis}
\label{section:risk_empirics}

Systemic credit risk is defined as that of cascade default of several banks triggered by default of one or several banks on its obligations \footnote{In the present study we restrict our analysis to cascades trigged by the default of a single bank.}. A default is operationally defined as an event where  the Capital Adequacy Ratio (CAR) defined below in Eq.~(\ref{eqCAR1}) falls below the minimal threshold defined by the regulator, the Central Bank of Russian Federation\footnote{This is also a legal reason for the Bank of Russia to revoke a license.}. The general definition of CAR reads \cite{CBRF_2004}:
\begin{equation}
\begin{split}
CAR=\frac{K-\sum\limits_{i=1}^{5}P_i}{\sum\limits_{i=1}^{5}(A_i-P_i)\cdot RW_i+O},
\end{split}
\label{eqCAR1}
\end{equation}
where K stands for the capital, $A_i$ denotes the $i$-th group of assets\footnote{The regulator breaks all assets into 5 types \cite{CBRF_market}.}, $RW_i$ is the corresponding risk weight and $P_i$ the corresponding  provision for non-performing loans specified by the regulator while $O$  is a collective notation for other risk variables such as market and operational risks which also used under CAR calculation. As the present analysis concentrates on risks related to interbank loans for which the risk weight $RW_{\rm IC}$ is equal to 20\%, in what follows we shall use the following simplified version of Eq.~(\ref{eqCAR1}):
\begin{equation}
\begin{split}
CAR=\frac{K-P_{\rm IC}}{0.2\cdot (A_{\rm IC}-P_{\rm IC})+{\tilde O}},
\end{split}
\label{eqCAR2}
\end{equation}
where $A_{\rm IC}$ and $P_{IC}$ denote interbank claims and corresponding regulator specified provision respectively  and $\tilde{O}$ denotes risk variables additional to those characterizing interbank credit obligations.

After a counterparty defaults on its obligations its lenders, in accordance with Russian regulatory document on rules for provision forming, have to form loan loss provision on deals with this counterparty.  According to Russian legislation banks have to establish provisions in accordance with borrower's quality and quality of debt servicing. In our simulations we will assume a provision of 100\% and take into account only provision for interbank deposit market operations considering the volume of other operations as fixed. 

To assess systemic risk we calculate the size of default cluster triggered by the bankruptcy of a particular bank. The stress-testing procedure we use is as follows:
\begin{itemize}
\item A default of a particular bank is assumed. All its creditors form provision on deals with this default counterparty and recalculate their CAR. 
\item We check whether the new CARs meet regulatory minimum (10\% for deposit taking banks which are allowed to attract deposit from individuals and 12\% for other non-banking activities like depositary, settlement and payment).  
\item The procedure is repeated for those creditors for which their CARs fall below the regulatory minimum. 
\end{itemize}

\begin{figure}[h!]
\begin{minipage}[h]{0.49\linewidth}
\centering
\includegraphics[width=\textwidth]{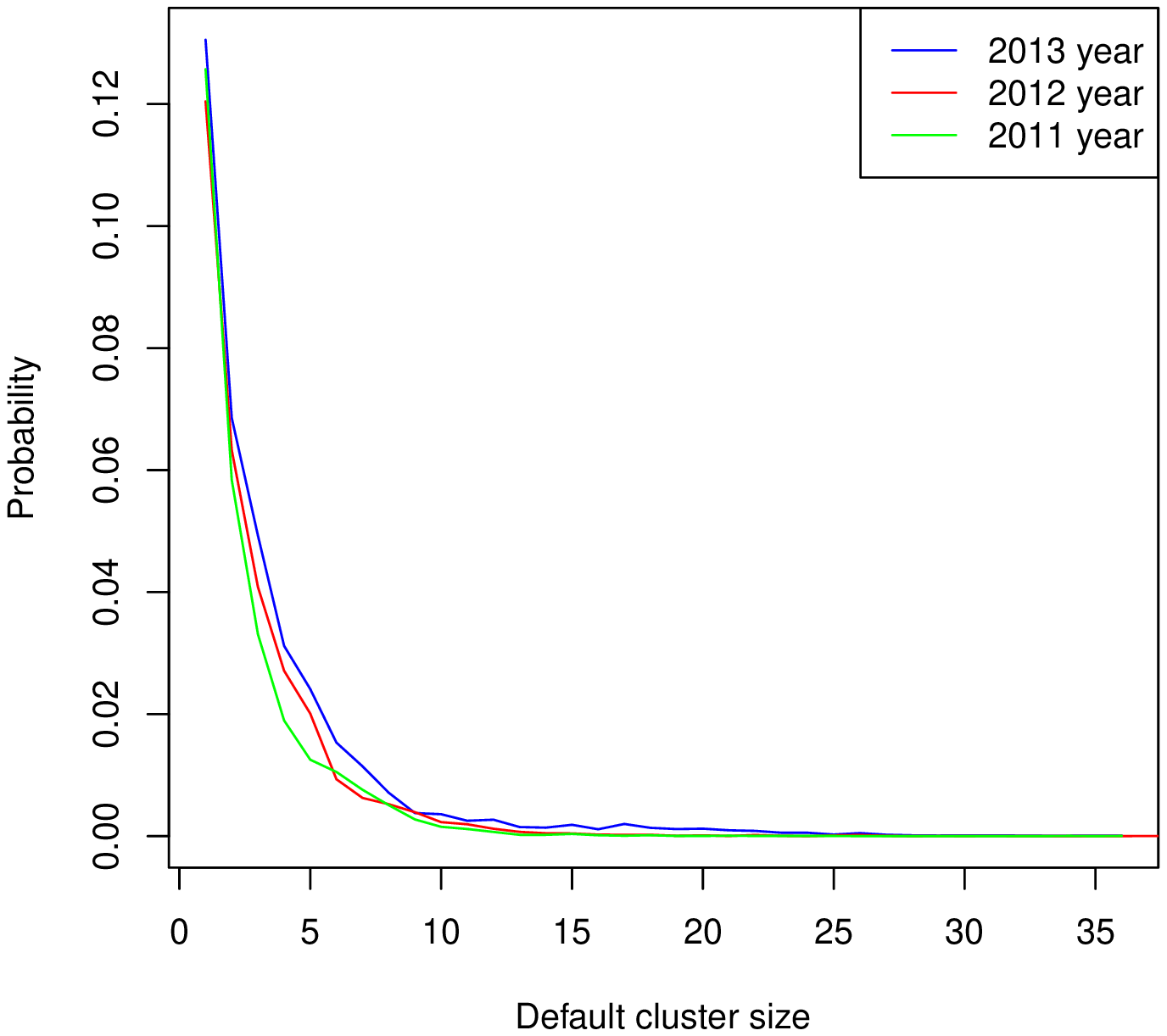}
\caption{Probability distribution for default cluster size}
\label{FDefClDistr}
\end{minipage}
\hspace{0.5cm}
\begin{minipage}[h]{0.49\linewidth}
\centering
\includegraphics[width=\textwidth]{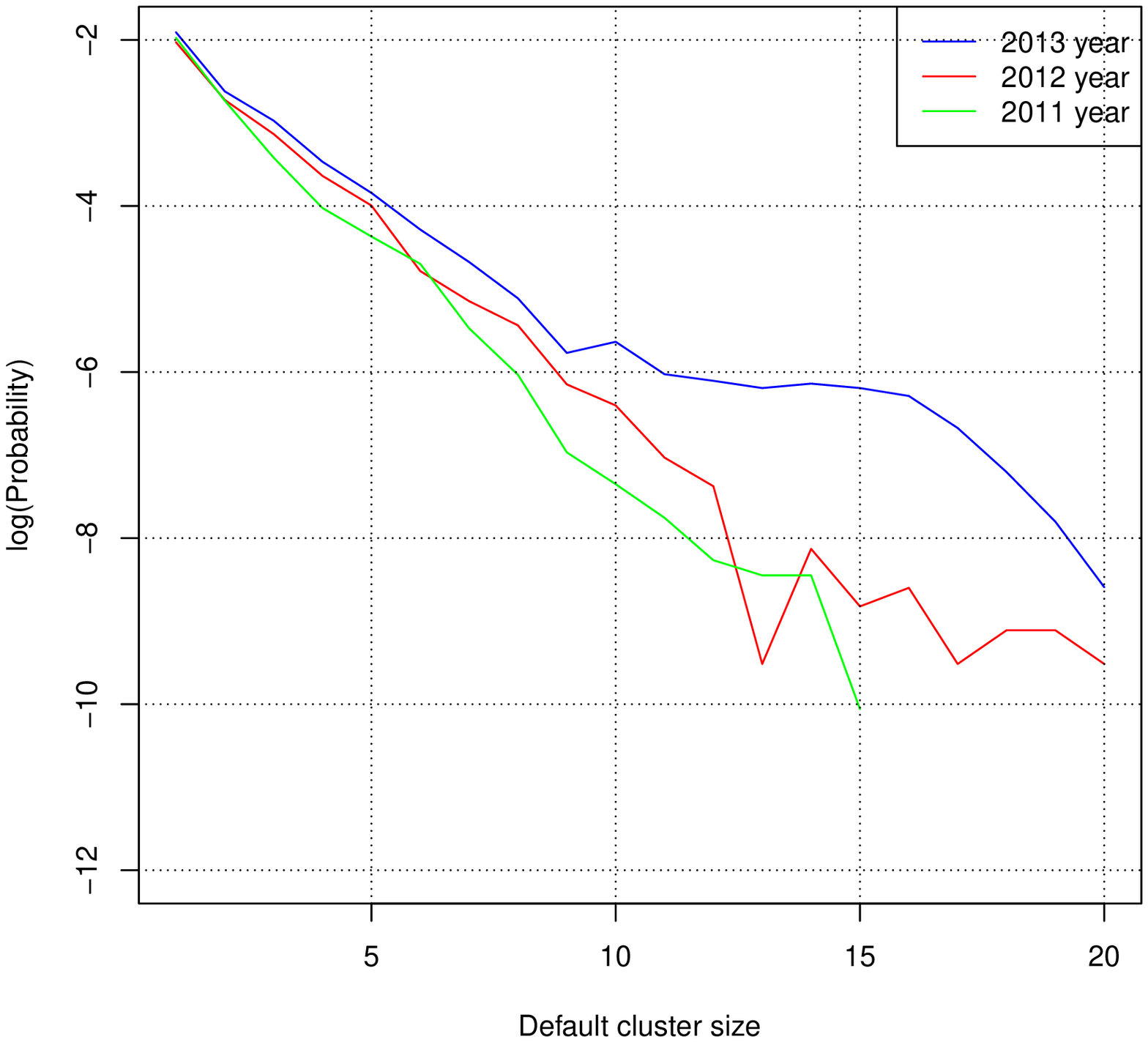}
\caption{Probability distribution for default cluster size, linear-log scale}
\label{FLogDefClDistr}
\end{minipage}
\end{figure}

The procedure is repeated for each bank from Out- and In-Out- components\footnote{It is clear that for the type of contagion  under study banks belonging to empty and In- components do not generate any systemic risk.}. The resulting probability distribution over the size of default clusters, where size is defined as a number of banks defaulting as a consequence of the default of an initial default node,  is shown, on the annual basis, in Table~\ref{tab:cluster_distribution} and Figs.~\ref{FDefClDistr} , \ref{FLogDefClDistr} (in the latter - on the log-linear scale).
\begin{table}[h!]
\begin{tabular}{|c|c|c|c|c|c|c|c|c|c|c|c|c|c|c|} \hline
Yr $\backslash$ $S_d$ & 1 & 2 & 3 & 4 & 5 & 6 & 7 & 8 & 9 & 10 & 11 & 12 & 13 & 14\\ \hline
2011 & 13.9 & 6.5 & 3.3 & 1.8 & 1.3 & 0.9 & 0.4 & 0.2 & 0.1 & 0.1 & 0 & 0 & 0 & 0\\ \hline
2012 & 13.2 & 6.6 & 4.4 & 2.6 & 1.8 & 0.8 & 0.6 & 0.4 & 0.2 & 0.2 & 0.1 & 0.1 & 0 & 0\\ \hline
2013 & 14.8 & 7.3 & 5.1 & 3.1 & 2.1 & 1.4 & 0.9 & 0.6 & 0.3 & 0.4 & 0.2 & 0.2 & 0.2 & 0.2\\ \hline
\end{tabular}
\caption{Probability distributions (in percent) for default cluster sizes in 2011, 2012 and 2013. For convenience the point corresponding to the 
probability of zero-sized cluster was removed}
\label{tab:cluster_distribution}
\end{table}
A first important conclusion that can be drawn from the Table~\ref{tab:cluster_distribution} and Figs.~(\ref{FDefClDistr},\ref{FLogDefClDistr}) is that the distribution over the size of default clusters is approximately exponential. This is a natural consequence of the (approximately) exponential degree distribution of the nodes belonging to default clusters (not shown), see e.g. \cite{Newman_2007}. 


Another important feature revealed by stress-testing is a higher importance in terms of generating default cascades and, therefore, systemic risk, of the banks from In-Out component as compared with those from the Out- one. In Table \ref{T_Def_Prob} we show the percentage of cases in which a default of the node under consideration triggers a default of another bank for the In-Out- and In- components ($29-38\%$ and $6-10\%$ respectively). The contribution of the In-Out- component is clearly the dominant one so that in what follows we will neglect the contribution of the Out- component. 
\begin{table}[h!]
\begin{center}
\begin{tabular}{|c|c|c|c|} \hline
Component $\backslash$ Year & 2011 & 2012 & 2013\\ \hline
In-Out & 0.29 & 0.31 & 0.38\\ \hline
Out & 0.06 & 0.10 & 0.10\\ \hline
CAR & 14.7 & 13.7 & 13.5\\ \hline
\end{tabular}
\caption{Bow-tie breakup of percentage of cascades and CAR's for Russian interbank market in 2011-2013}\label{T_Def_Prob}
\end{center}
\end{table}

\begin{figure}[h!]
\centering
\includegraphics[width=0.8\linewidth]{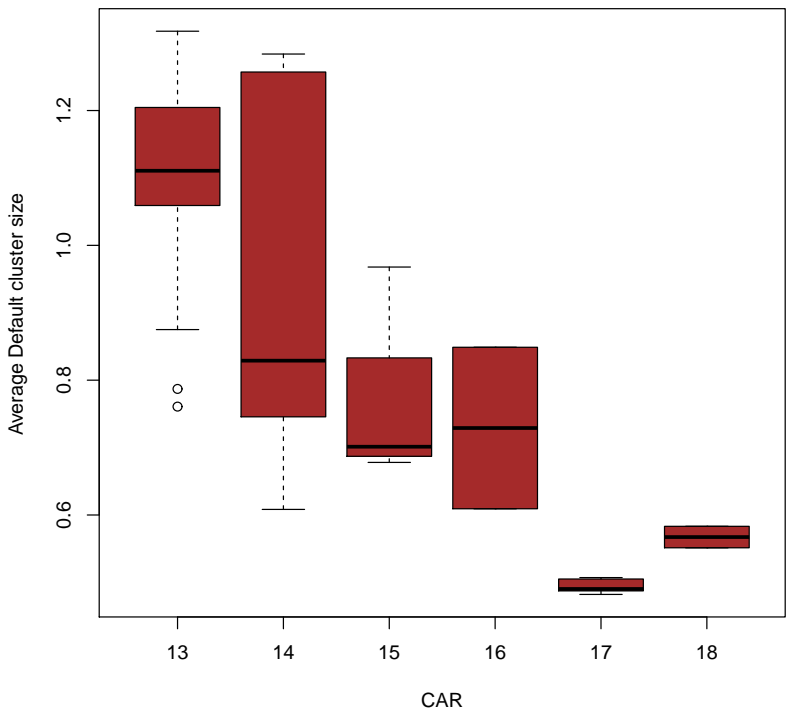}
\caption{Dependence of default cluster size on CAR}
\label{SCAR}
\end{figure}

From Table \ref{T_Def_Prob} we also see that the average stability of banks with respect to default risks as characterized by the average CAR underwent, between 2011 and 2013, a significant reduction. It is quite clear that lower values of CAR generate larger systemic risks. Indeed, an analysis in \cite{Kapadia_2010} has shown a dramatic dependence of contagion on capital reserves. A dependence of the average size of default cluster on CAR (calculated on the monthly basis) is shown in Fig.~\ref{SCAR}\footnote{The bottom and top of the box are the first and third quartiles and the band inside the box is the median. The end of the low whisker is the lowest datum still within 1,5 x interquartile range and the highest datum still within 1,5 x interquartile range of the upper quartile.”}.  
From Fig.~\ref{SCAR} it is clearly seen that there indeed exist a pronounced dependence of the magnitude of contagion on the capital reserves reflected by CAR. 

\begin{figure}[h!]
\centering
\includegraphics[width=\textwidth]{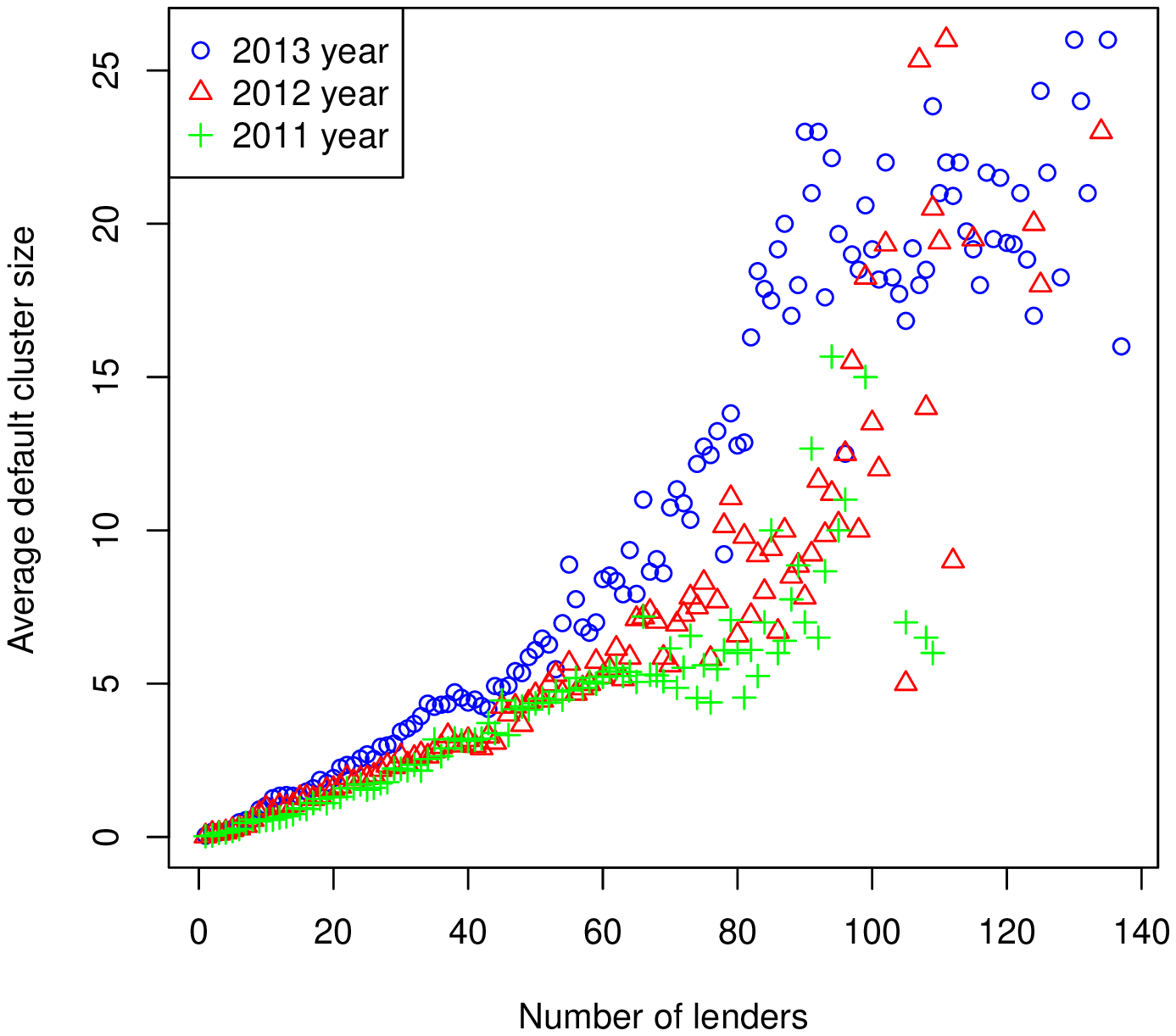}
\caption{Dependence of default cluster size on the number of lenders}
\label{FDefClOnLink}
\end{figure}

It is quite typical for contagion that its volume grows with increasing centrality of the source node, see e.g. \cite{Borge_2013}. In Fig.~\ref{FDefClOnLink} we plot a dependence of the average size of the default cluster on the out-degree of the bankrupt node (i.e. the number of lenders), where for each day and each bank we calculate the number of lenders and the size of default cluster generated by this bank at this day and average over all days in a given year. 
We see that indeed the volume of contagion increases with increasing out- degree and that for large out-degrees the dependence in question is distinctly nonlinear so that the volume of contagion shows a faster than linear growth with the out-degree of the source node. From Fig.~\ref{FDefClOnLink} one can also conclude that, following the above-described reduction in CAR from 2011 to 2013, a volume of contagion has dramatically grown within this period.  

\begin{figure}[h!]
\begin{minipage}[h]{0.49\linewidth}
\centering
\includegraphics[width=\textwidth]{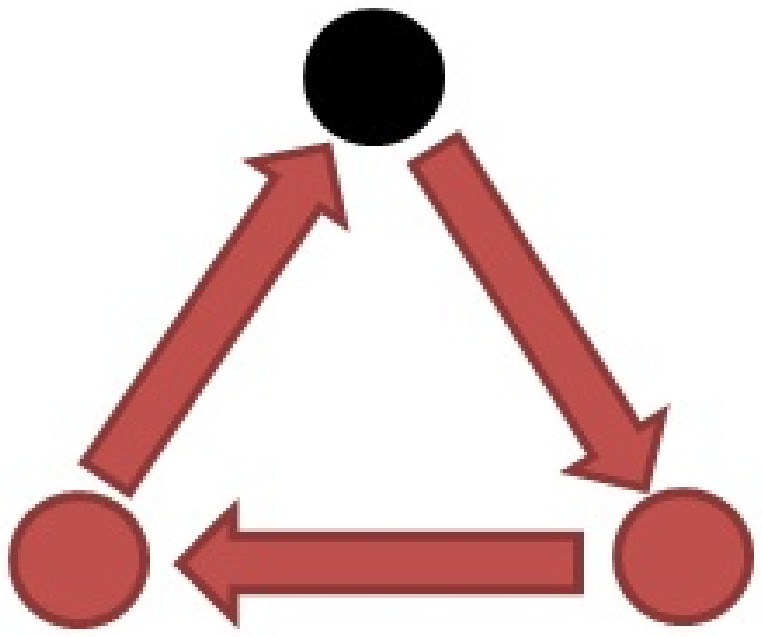}
\caption{Motifs of length 3 in default cluster, type T1}
\label{FExampleMotifA}
\end{minipage}
\hspace{0.5cm}
\begin{minipage}[h]{0.49\linewidth}
\centering
\includegraphics[width=\textwidth]{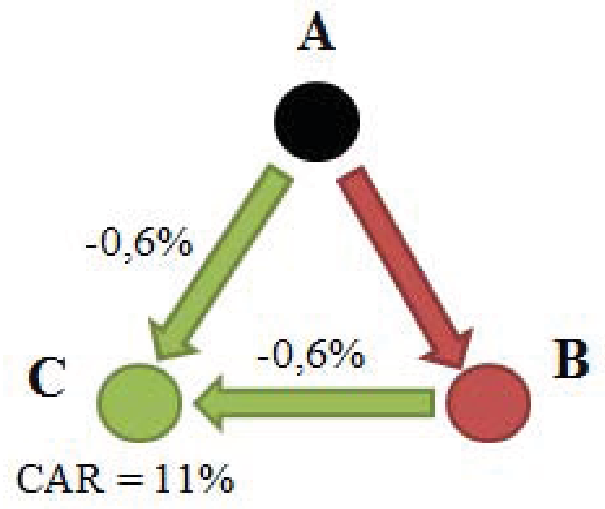}
\caption{Motifs of length 3 in default cluster, type T2}
\label{FExampleMotifB}
\end{minipage}
\end{figure}

The key question in developing a model for propagation of contagion is that of topology of default clusters. In our simulation we found out that default clusters combining vulnerable banks are, with very few exceptions, tree-like with the maximal length of branches equal to 4. In directed graphs the simplest nontrivial motifs, triangles, can belong to two types, T1 and T2, differing by orientation of participating links, see Figs. \ref{FExampleMotifA} and  \ref{FExampleMotifB} correspondingly.
Let us consider for definiteness an example shown in Fig~\ref{FExampleMotifB}, where the starting event is a default of the bank A leading to f critical CAR loss of B, but leaving C with an admissible CAR of 10.4 \% (with the regulatory minimum set at 10 \%). The consequent CAR loss of 0.6 \% from the link $B \to C$ does, however lead to the default of C.  Our simulations have shown that, over all the days,  the average number of triangles of the type T1 is 0.06 and of the type T2 -  1.2. Both numbers are small as compared to total number of default clusters per day considered in our simulations. The effectively tree-like structure of the default cluster that allows to neglect motifs in working out a mathematical model for default cascading. In particular, under this assumption propagation of vulnerability between the nodes can happen only along a single link. 


The last key ingredient for modelling default propagation is to quantify vulnerability of a given node with respect to default of at least one of its neighbors. Vulnerability can most naturally be described as a probabilistic characteristic of link's ability to transport contagion from one node to another. We call a link vulnerable if a default of the counterparty may lead to default of another counterparty through this link. The link vulnerability depends on the local geometry of a network. Generically it is defined as a conditional probability $v(r,s \vert k,l)$ of default propagation from the node with in- and out-degrees $r,s$ to the node with in- and out- degrees $k,l$. In addition vulnerability depends on the position of the two nodes under consideration within the bow-tie structure of the network so that probabilistic pattern of contagion propagation is specified by the conditional probabilities $v^{\rm IO \to IO} (r,s \vert k,l)$, $v^{\rm IO \to In} (r,s \vert k,l)$ and their more complex modifications. The corresponding empirical vulnerability distributions are found to differ a lot, see Refs~\cite{Leonidov_2012,Leonidov_2014}.

\section{Systemic credit risk. Mathematical modeling}
\label{section:risk_theory}

The mathematical model we use to describe systemic risks on the Russian interbank market is based on empirical findings described above in the Sections~\ref{section:deposit}, \ref{section:risk_empirics}. Let us reiterate the main features of importance for description of contagion process:
\begin{itemize}
\item Degree distributions are fat-tailed.
\item The network is characterized by significant disassortative correlations between adjacent nodes. 
\item Because of significant differences in systemic risks related to the position of the infected node and its neighbors within the bow-tie structure it is natural to take this into account explicitly when building the mathematical formalism\footnote{Although the mathematical construction built in the present paper is quite natural and is versatile enough to describe the empirical simulations of systemic risk, the question of whether it can be simplified further is relevant and deserves further investigation. An interesting example of solving the problem of this kind can be found in Ref.~\cite{Fagiolo_2013}. }.
\item Default clusters are tree-like.
\end{itemize}
 
\begin{figure}[h]
\begin{minipage}[h]{0.95\linewidth}
\centering
\includegraphics[width=\textwidth]{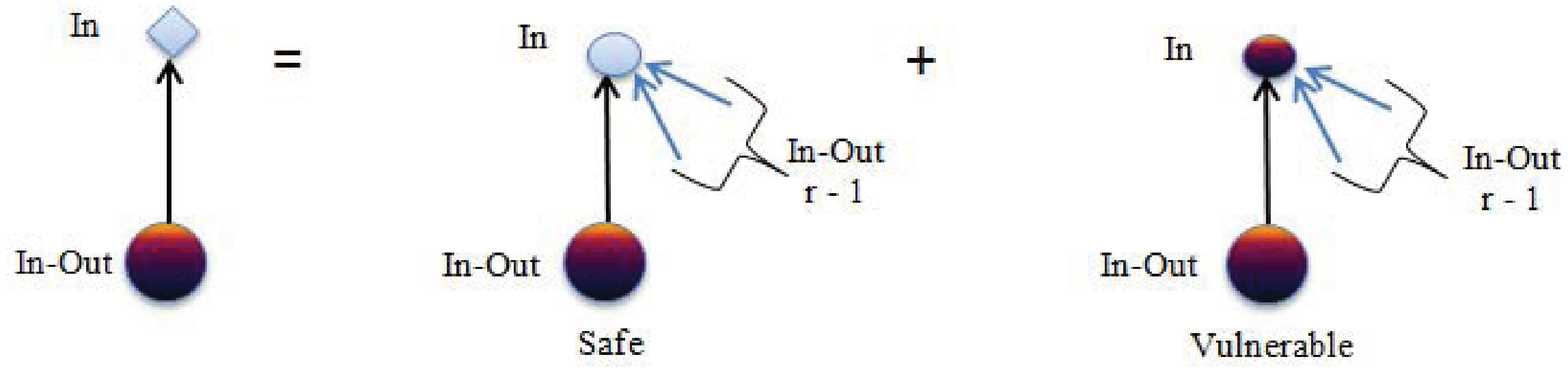} a) \\
\end{minipage}

\vfill

\begin{minipage}[h]{0.95\linewidth}
\centering
\includegraphics[width=\textwidth]{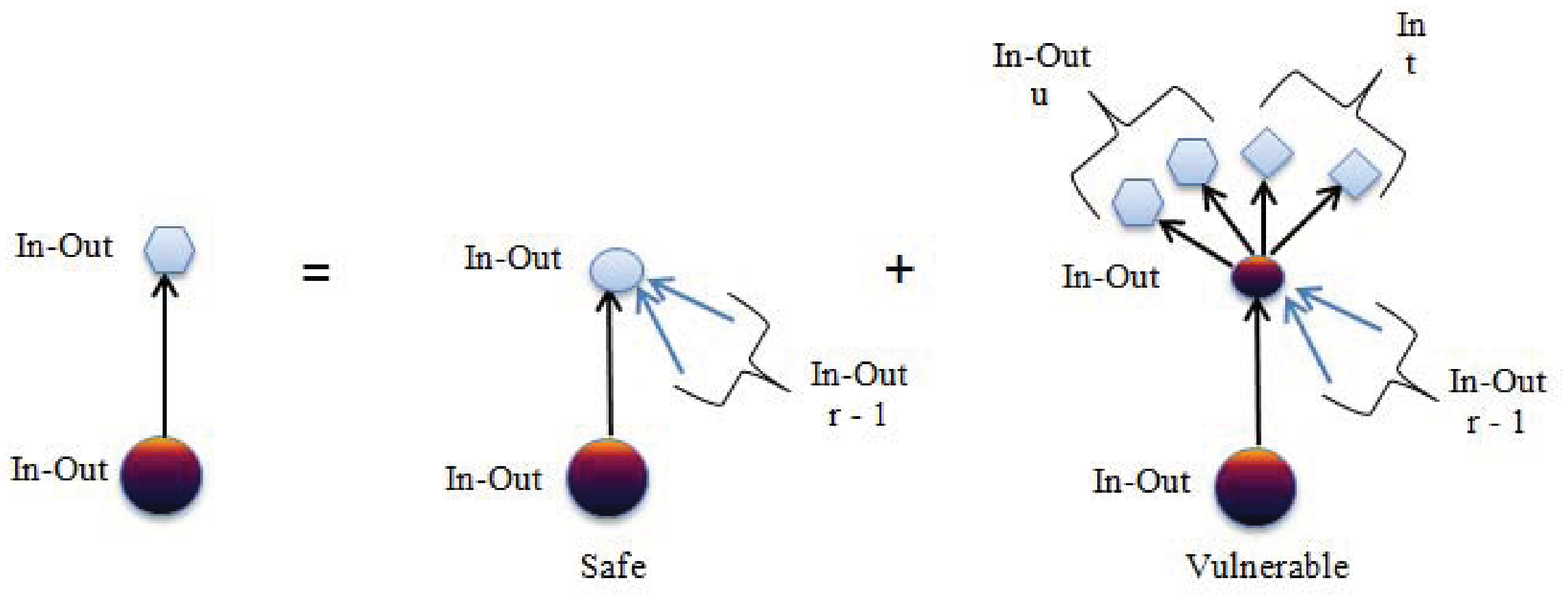} b) \\
\end{minipage}
\caption{Default spreading mechanism. The individual diagrams depict corresponding probabilities, see e.g. similar notation in \cite{Kapadia_2010}.}
\label{FDefSpr}
\end{figure}

Let us consider a bank from the In-Out component with $k+l$ outgoing links, where $k$ of them lead to the In-Out- component and $l$ to In- component respectively\footnote{As discussed above, nodes from the Out- component generate very small systemic risks so that the corresponding effects will be neglected} and take a randomly chosen edge linking the chosen node to a node in the In- component which, in addition, has $r-1$ incoming links, see Fig. \ref{FDefSpr} a. This is a simplest case where contagion goes from the In-Out- component to the In-one and stops there. 

Conditional probability distribution for reaching a vulnerable bank from the In- component following a link from the In-Out- one is described by the generation function $N_{k,l}(y)$
\begin{equation}
N_{k,l}(y)=\sum\limits_{r}^{\infty} P^{\rm IO \to In}(r|k,l)\left(1-v^{\rm IO \to In}(r|k,l)+y\;v^{\rm IO \to In}(r|k,l) \right)
\label{eqGenFuncInOut/In}
\end{equation}
where we have taken into account that an outgoing link under consideration can lead to a vulnerable bank with probability $v^{\rm IO \to In}(r|k,l)$ or safe one with probability $1-v^{\rm IO \to In}(r|k,l)$ and, through the conditional probability $P^{\rm IO \to In}(r|k,l)$, the probabilistic interdependence of the degrees of nodes connected by this link. The corresponding part of the default cluster can therefore described as a projection of the initial network onto a graph in which a link in the original graph survives with the probability  $P^{\rm IO \to In}(r|k,l)\;v^{\rm IO \to In}(r|k,l)$. 

Let us now consider a randomly chosen edge linking two banks from the In-Out- component, see Fig.~\ref{FDefSpr} b. The equation for the corresponding generating function $M_{k,l}(x,y)$ reads\footnote{The notations for the indices should be clear from Fig.~\ref{FDefSpr} b.}: 
\begin{equation}
\begin{split}
M_{k,l}(x,y)=\sum\limits_{u,t,r}^{\infty} P^{\rm IO \to IO}(u,t,r|k,l)(1-v^{\rm IO \to IO}(u,t,r|k,l))+\\
x\sum\limits_{u,t,r}^{\infty} P^{\rm IO \to IO}(u,t,r|k,l) v^{\rm IO \to IO}(u,t,r|k,l)[M_{u,t}(x,y)]^{u}[N_{u,t}(y)]^{t}
\end{split}
\label{eqGenFuncInOut/InOut}
\end{equation}
The corresponding part of the default cluster can be described as a projection of the initial network onto a graph in which a link in the original graph survives with the probability  $P^{\rm IO \to IO}(u,t,r|k,l)\;v^{\rm IO \to IO}(u,t,r|k,l)$. 

Let us define a generation function $F(x,y)=\sum\limits_{k,l}^{\infty} P^{\rm IO}(k,l)x^ky^l$ for the probability for a bank from the In-Out- component to have $k$ and $l$ first neighbors from the In-Out- and In- components respectively. Then the generation function for the number of vulnerable banks in the network is simply
\begin{equation}
F \left( \{ M_{kl} (x,y) \}, \{N_{kl} (y) \} \right) \equiv F(M,N) = \sum\limits_{k,l}^{\infty} P^{\rm IO}(k,l)\left[ M_{kl}(x,y) \right]^k \left[ N_{kl}(y) \right]^l
\end{equation}
It is easy to see that for calculation of the mean default cluster size $S$ one can put $y=x$ and compute a derivative of $F(M,N)$ at point $x=1$. We have
\begin{equation}
S dx = dF(M,N)_{y=x=1}=\sum\limits_{k,l}^{\infty} P^{\rm IO}(k,l)(k dM_{k,l | x=1}+ldN_{k,l | x=1}),
\label{eqDiff0}
\end{equation}
where we have used the normalization property of generation functions \\ $M_{k,l|x=1}=N_{k,l|x=1}=1$. From Eq.~(\ref{eqGenFuncInOut/In}) we get 
\begin{equation}
dN_{k,l|x=1}=\sum\limits_{r}^{\infty} P^{\rm IO \to In}(r|k,l)v^{\rm IO \to In}(r|k,l)dx, 
\label{eqDiffN}
\end{equation}
so that
\begin{equation}
dM_{k,l}=\sum\limits_{u,t}^{\infty} \alpha_{u,t,k,l} dM_{u,t}+\gamma_{k,l}dx
\label{eqDiffM}
\end{equation}
where 
\begin{equation}
\alpha_{u,t,k,l}=\sum\limits_{r}^{\infty} u P^{\rm IO \to IO}(u,t,r|k,l)v^{\rm IO \to IO}(u,t,r|k,l)
\end{equation}
\begin{eqnarray}
\gamma_{k,l} & = & \sum\limits_{u,t,r}^{\infty} P^{\rm IO \to IO}(u,t,r|k,l)v^{IO \to IO}(u,t,r|k,l) \nonumber \\
& + & \sum\limits_{u,t,r}^{\infty} P^{IO \to IO}(u,t,r|k,l)v^{\rm IO \to IO}(u,t,r|k,l) \nonumber \\
& \times & t \; \sum\limits_{r1}^{\infty} P^{IO \to In}(r_1|u,t)v^{\rm IO \to In}(r_1|u,t)
\end{eqnarray}
It is useful to rewrite Eq.~(\ref{eqDiffM}) in the operator form:
\begin{equation}
dM=A dM+\gamma dx,
\label{eqDiffMOp}
\end{equation}
where $dM$ and $\gamma$ are vectors of length $k \times l$ and $A$ is a $k \times l,k \times l$ matrix  size with the elements $A_{(k,l)(u,t)}=\alpha_{u,t,k,l}$. For solution of Eq.(\ref{eqDiffMOp}) to exist its maximal eigenvalue $\lambda_{max}$ should satisfy $\lambda_{max}<1$\footnote{In \cite{Leonidov_2014} it is shown that a more restrictive condition valid in the non-percolative regime reads $\sum\limits_{u,t}^{\infty}A_{(k,l)(u,t)}<1$ for all pairs $(k,l)$.}. It reads:
\begin{equation}
dM_{k,l}=\sum\limits_{u,t}^{\infty} \beta_{k,l,u,t} \gamma_{u,t}dx
\label{eqDiffMSol}
\end{equation}
where $\beta_{u,t,k,l}$ is an element $B_{(u,t),(k,l)}$ of the matrix $B=(I-A)^{-1}$. Equation~(\ref{eqDiffMSol}) is valid in the absence of a giant cluster and should be modified in the percolative phase, see e.g. \cite{Newman_2002,Boccaletti_2006}.

\begin{figure}[h!]
\centering
\includegraphics[width=0.9\linewidth]{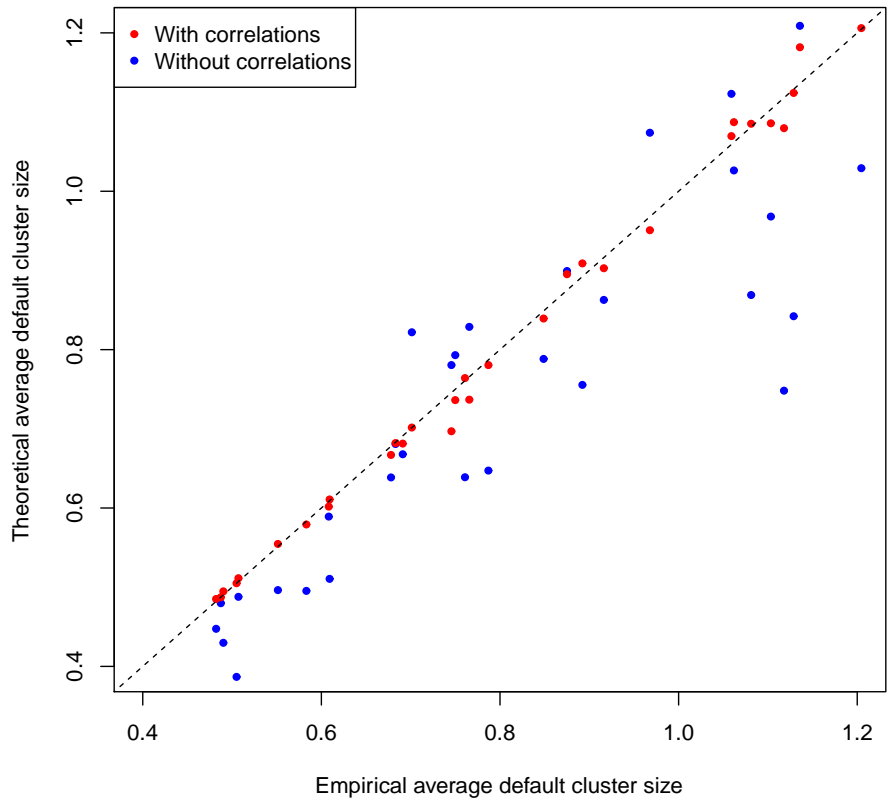}
\caption{Comparison of theoretical and empirical average default cluster size with and without accounting for degree-degree correlations}
\label{FDefClusterSizeCompare}
\end{figure}

The final equation for the average default cluster size $S$ following from Eqs.~(\ref{eqDiff0},\ref{eqDiffN},\ref{eqDiffM},\ref{eqDiffMOp}) reads
\begin{equation}
S  = \left. \frac{dF}{dx} \right \vert_{x=1}=\sum_{k,l}^{\infty}P^{IO}(k,l) \left[ k\sum_{u,t}^{\infty}\beta_{k,l,u,t}\gamma_{u,t}+l\sum_{r}^{\infty}\omega_{k,l,r} \right], 
\label{eqS}
\end{equation}
where
\begin{equation}
\omega_{k,l,r}=P^{\rm IO\rightarrow In}(r|k,l)\upsilon^{\rm IO\rightarrow In}(r|k,l).
\end{equation}
To calculate the average size of the default cluster we use empirical conditional probability distributions  $P^{\rm IO \to In}(r|k,l)$, $v^{\rm IO \to In}(r|k,l)$, $P^{\rm IO \to IO}(u,t,r|k,l)$ and $v^{\rm IO \to IO}(u,t,r|k,l)$ that are calculated on the monthly basis. A comparison of the model predictions and results of stress testing for the average default cluster size is shown in Fig.\ref{FDefClusterSizeCompare}. We see a very good agreement between the model and experiment provided one takes into account correlations between the degrees of adjacent nodes captured by $P^{\rm IO \to In}(r|k,l)$ and $P^{\rm IO \to IO}(u,t,r|k,l)$ and a much poorer one when these correlations are neglected. The remaining deviations can be ascribed to using analytical approximation appropriate to infinite graphs\footnote{As shown in section \ref{section:risk_empirics} the maximal number of branchings in default trees is 4.}  Another source of deviation is in neglecting triangles in default graph. 

\section{Conclusions}
\label{section:conclusions}

Let us formulate once again the main conclusions of the paper.

In this study we have described a new analytical model of default contagion propagation taking into account the difference in the functional role of the nodes (bow-tie decomposition), realistic in- and out- degree and link weight distributions and disassortative degree-degree correlations. Its predictions are shown to be in line with the results of Monte-Carlo simulations.

Let us reiterate, that to build a successful model of contagion propagation in interbank networks one needs to combine empirical studies and adequate theoretical framework. Empirical information of importance is related both to topological characteristics of the interbank market network under consideration and to the process of contagion propagation from node to node that depends, in particular, on interplay between link and node characteristics (volume of loans and bank balance sheets respectively). It was shown that very good description of default cascade simulation can be given in the formalism that explicitly takes into account degree distributions and degree-degree correlations and conditional probabilities of contagion propagation from one node to another.  Let us mention that although a marked difference in conditional default probabilities for nodes belonging to different components and empirical simulation results highlighting the dominant role of the in-out component provide strong arguments for explicit account of the bow-tie structure in describing systemic contagion risks, a more detailed analysis of this issue (in particular going beyond considering the mean size of default clusters) is certainly desirable. We are planning to address this problem in a separate publication.
The results obtained in this paper can be used for estimating systemic risks in interbank networks as well as for analyzing sensitivity of systemic risk with respect to changes in network topology and stability of individual banks. The results can also be used for the analysis of liquidity risks -- with the contagion propagation through in-links and modification of vulnerability criterion. 

\bigskip

{\bf Acknowledgements}

\medskip

We are grateful to the referees for the comments that helped to clarify the presentation of our results.

\end{document}